\documentstyle[12pt]{article}
\begin{document}
\baselineskip = 20pt

\title{GENERALIZED THERMAL ZETA-FUNCTIONS}

\author{ H. Boschi-Filho$^{\star}$ and C. Farina$^{\dagger}$\\
\\
Instituto de F\'\i sica - Universidade Federal do Rio de Janeiro \\
Cidade Universit\'aria - Ilha do fund\~ao - Caixa Postal 68528 \\
21945-970 Rio de Janeiro, BRAZIL.}

\bigskip
\maketitle
\begin{abstract}
We calculate  the partition function of a harmonic oscillator with
quasi-periodic boundary conditions using the zeta-function method.
This work generalizes a previous one by Gibbons and contains the
usual bosonic and fermionic oscillators as particular cases.  We give
an alternative prescription for the analytic extension of the
generalized Epstein function involved in the calculation of the
generalized thermal zeta-functions. We also conjecture about the
relation of our calculation to anyonic systems.
\end{abstract}
\bigskip
\vfill
\noindent $^\star${e-mail: boschi@if.ufrj.br}\par
\noindent $^\dagger${e-mail: farina@if.ufrj.br}
\par

\pagebreak

Since the generalized zeta-function method for computing determinants
was introduced into theoretical physics calculations
\cite{Salam}, this method has been applied successfully in different
areas of physics. Examples vary from ordinary quantum and statistical
mechanics to quantum field theory where anomalies, fermionic
determinants as well as all sorts of effective actions can be
computed by the generalized zeta-function prescription.

One of the first and very illustrative examples of this method,
published in this Journal by Gibbons \cite{Gibbons}, concerns the
computation of the partition functions of both the bosonic and
fermionic oscillators.  The application of the zeta-function method
for these cases demands, the analytical extension of the so called
inhomogeneous one-dimensional Epstein function (see for instance ref.
\cite{GV} for some mathematical details on Epstein functions).

However, as far as we know, the interpolating system between the
bosonic and fermionic oscillators has never been treated by the
generalized zeta-function method in the literature.  Recently, this
problem has been solved by the authors using the Green function
method \cite{BFD}.

In this letter, we generalize Gibbons work \cite{Gibbons} since we
compute $\det (\omega^2 -\partial_t^2)$ with a general boundary
condition, which contains the periodic and antiperiodic boundary
conditions as particular cases, by using the generalized
zeta-function prescription. To do that, an analytical extension of
what is called generalized inhomogeneous one-dimensional Epstein
function is needed.  We solve this problem generalizing the procedure
of Ambjorn and Wolfran \cite{AW} to the usual Epstein function and
obtaining this way a very convenient expression for the problem at
hand.

Let us then calculate the partition function for a harmonic
oscillator with quasi-periodic boundary conditions. For an usual
(bosonic) harmonic oscillator and even for a fermionic oscillator the
corresponding partition functions can be obtained by various methods,
in particular by a path integral continued to the Euclidean space

\begin{equation}
{\cal Z}(\beta) = \int [d\mu_{x}] e^{-S}\;,
\end{equation}

\noindent where $S$ is the classical action. For the bosonic case we
have $S=(1/2)\int_0^\beta x(\omega^2 - \partial_t^2) x dt$ and the
integral is taken over periodic functions on the interval
$(0,\beta)$:

\begin{equation}
x(t+\beta)=x(t)\;.
\end{equation}

The fermionic case can be treated on similar grounds substituting the
classical (c-number) functions $x$ by anticommuting Grassmann
variables $x$ and $\bar x$, which must be antiperiodic on the
interval $(0,\beta)$:

\begin{equation}
x (t+\beta) = - x (t) \;;\;\;\;\;\;\;
\bar x (t+\beta)= - \bar x (t) \;
\end{equation}

\noindent and the path integral measure is $[dx][d\bar x]$.

For the generalized case, however, there is no such a prescription.
In order to circumvent this difficulty we follow the work of Gibbons
who first applied the zeta-function method to the calculation of the
partition functions for the bosonic and fermionic oscillators as the
following determinants

\begin{equation}
{\cal Z}^{Bosonic} (\beta) =
\left. { \det}^{-1/2} (\omega^2 - \partial_t^2 ) \right|_{periodic}
\end{equation}

\noindent and

\begin{equation}
{\cal Z}^{Fermionic} (\beta) =
\left. { \det}^{+1} (\omega^2 - \partial_t^2 )\right|_{antiper.}  \;.
\end{equation}

As our starting point, we shall then take as the definition of the
partition function for the general case, the following determinant

\begin{equation}\label{Z}
{\cal Z}_\sigma^\theta (\beta) =
\left.{\det}^\sigma (\omega^2 - \partial_t^2 ) \right|_\theta
={ \det}^\sigma (L)_\theta\;,
\end{equation}

\noindent where $\sigma$ is a parameter related to statistics and
$\theta$ to the quasi-periodic boundary condition of the chosen
oscillator. By quasi-periodic boundary condition we mean that

\begin{equation}\label{cond}
x(t+\beta)= e^{i\theta} x(t)\;.
\end{equation}

\noindent
The particular cases of the bosonic and fermionic
oscillators can be obtained by taking $\sigma=-1/2$, $\theta=0$ and
$\sigma=+1$, $\theta=\pi$, respectively, as we shall show bellow.

The eigenvalues of the operator $L=\omega^2 - \partial_t^2$ under the
quasi-periodic boundary condition are:

\begin{equation}
\lambda_{m}^\theta = \omega^2 +
\left({\theta+2m\pi\over\beta}\right)^2 \;;\;\;\;\;\;\;\;\;\;
(m=0, \pm 1, \pm 2, \dots)
\end{equation}

\noindent and the corresponding generalized zeta-function is given by

\begin{eqnarray}
\zeta^\theta (s;L)
& \equiv & \sum_m \left(\lambda_{m}^\theta\right)^{-s}\nonumber \\ &
= & \sum_{m=-\infty}^{+\infty} \left[ \omega^2 +
\left({\theta+2m\pi\over\beta}\right)^2 \right]^{-s} \nonumber \\
& = & \left({2\pi\over\beta}\right)^{-2s} {\cal D} (s,\nu,
{\theta\over 2\pi}) \;,\label{zeta}
\end{eqnarray}

\noindent where $\nu= \omega\beta /2\pi$ and we defined

\begin{equation}\label{D}
{\cal D} (s,\nu, {\theta\over 2\pi}) = \sum_{m=-\infty}^{+\infty}
\left[ \nu^2 +
\left({\theta\over 2\pi}+m \right)^2 \right]^{-s} \;.
\end{equation}

\noindent The function $\zeta^\theta (s;L)$ is the generalized
thermal zeta-function, which reduces to Gibbons cases for particular
values of $\theta$. From this function one can calculate the desired
partition function

\begin{equation}\label{Z2}
{\cal Z}_\sigma^\theta (\beta) = \exp
\left. \left\{ -\sigma {\partial\over\partial s} \zeta^\theta
(s;L) \right\} \right|_{s=0} \;.
\end{equation}

Once, the function defined by eq. (\ref{D}) is not analytic at the
origin we will have to make an analytic continuation for ${\cal D}
(s,\nu, {\theta\over 2\pi})$ in order to determine the partition
function ${\cal Z}_\sigma^\theta (\beta)$.  In fact, the application
of the generalized zeta-function method consists basically of the
following three main steps: first find the eigenvalues and construct
the series $\zeta (s) = \sum_m \lambda_{m}^{-s}$; then make an
analytical extension for the whole complex $s$-plane (or at least
into a region containing the origin) and finally compute $\exp
\left. \left\{ - {\partial\over\partial s} \zeta (s)
\right\}\right|_{s=0}$. For the case at hand, the function ${\cal D}
(s,\nu, {\theta\over 2\pi})$ is in fact related to the
one-dimensional generalized Epstein function $Z_1^{c^2} (s;1,a)$,
since

\begin{equation}\label{E}
Z_1^{c^2} (s;1,a) = \sum_{n=1}^{+\infty} \left[ c^2 + (n+a)^2
\right]^{-s} \;,
\end{equation}

\noindent for which there are well known analytic continuations
\cite{Epstein}. However, these formulas are not easy to handle and we
performed a new representation for ${\cal D} (s,\nu, {\theta\over
2\pi})$ which can be considered as a generalization of the one given
by Ambjorn and Wolfran \cite{AW}. In our case we obtain (see the
appendix for details):

\begin{equation}\label{ca}
{\cal D} (s,\nu,{\theta\over 2\pi}) = {\sqrt{\pi}\over \Gamma (s)}
\left[ {\Gamma (s-{1\over 2})\over
\nu^{2s-1}} + 4 \sum_{m=1}^{+\infty} \cos(m\theta) \left(
{m\pi\over\nu}\right)^{s-{1\over 2}} K_{s-{1\over 2}}
(2m\pi\nu)\right]\;,
\end{equation}

\noindent where $K_\alpha (x)$ is the modified Bessel function of the
second kind with order $\alpha$. From expression (\ref{ca}) it can
also be shown that (see the appendix)

\begin{equation}
{\cal D} (0,\nu,{\theta\over 2\pi}) = 0
\end{equation}
\noindent and
\begin{equation}\label{dD}
\left. {\partial\over\partial s} {\cal D} (s,\nu,{\theta\over 2\pi})
\right|_{s=0} = - \ln \left[ 2 \left( \cosh 2\pi\nu - \cos
\theta\right)\right]\;.
\end{equation}

\noindent
Substituting (\ref{dD}) into  (\ref{Z2}), we have:

\begin{eqnarray}
{\cal Z}_\sigma^\theta (\beta) & = & \left. \exp \left[ - \sigma
{\partial\over\partial s} {\cal D} (s,\nu,{\theta\over 2\pi}) \right]
\right|_{s=0} \nonumber\\ & = & \exp \left\{ \sigma \ln \left[ 2
\left( \cosh 2\pi\nu - \cos
\theta\right) \right] \right\} \nonumber\\
& = & 4^\sigma \left[ \cosh^2 {\omega\beta\over 2} - \cos^2
{\theta\over 2} \right]^\sigma\;.\label{Z3}
\end{eqnarray}

\noindent
This partition function will reduce naturally to the corresponding
one of the usual bosonic oscillator if we take $\sigma=-1/2$ and
$\theta=0$ \cite{Gibbons},

\begin{equation}
{\cal Z}^{Bosonic} (\beta) = \left( 2 \sinh {\omega\beta\over 2}
\right)^{-1} \;.
\end{equation}

\noindent Analogously, for the quadratic fermionic oscillator of
Finkelstein and Villasante \cite{FV} we have, after choosing
$\sigma=+1,
\theta=\pi$, that

\begin{equation}
{\cal Z}^{Fermionic}_{Quadratic} (\beta) =  4 \cosh^2
{\omega\beta\over 2} \;.
\end{equation}

\noindent The linear fermionic oscillator \cite{Das} is obtained with
the choice $L\rightarrow L^{1/2}; \sigma=+1, \theta=\pi$

\begin{equation}
{\cal Z}^{Fermionic}_{Linear} (\beta) =  2 \cosh {\omega\beta\over
2}\;.
\end{equation}

In fact, because of the general properties of determinants, the above
choice could also be $\sigma=+1/2$, $\theta=\pi$ and $L$ unchanged.
Curiously, Gibbons was able to find this result for the linear
fermionic oscillator without a similar rescaling in $L$. This was
possible since he considered, from the beginning, the sum in the
zeta-function (\ref{zeta}) only over positive values of $m$. This
naturally falls into our result since this introduces a factor $1/2$
for ${\cal D} (s,\nu,{\theta\over 2\pi})$ and results in a
corresponding power for the determinant, as can be seen, for example,
in eq. (\ref{dD}).

A physical interpretation of our result is also possible. Equation
(\ref{Z3}) corresponds to the partition function of what we call the
anyonic oscillator, in the sense that, besides interpolating
continuously between the bosonic and fermionic cases it contains these
two cases as particular choices of the parameters involved.  This
result coincides with the one given by the authors using the Green
function method to calculate the ratio of two determinants
\cite{BFD}.

Actually, anyons are expected to live in two space dimensions
\cite{Wilczek}, but the generalized statistics they are related to,
can also be defined in one space dimension \cite{LM} as we have done
here. Furthermore, it has been shown in the literature that anyons
confined to the lowest Landau level correspond to anyons in one
dimension \cite{HLM}.

As a satisfactory quantization for anyonic systems are not known yet,
we believe that our result may contribute towards this direction.
Besides, we hope that our analytic continuation for ${\cal D}
(s,\nu,{\theta\over 2\pi})$ will be of help in discussing other
problems, specially in quantum field theory as the computation of
effective actions at finite temperature, the Casimir effect, etc.

\bigskip
\noindent
{\bf Acknowledgements}.  The authors were partially supported by CNPq,
Brazilian agency.

\bigskip

\font\title=cmb10 scaled\magstep2
\centerline{\title Appendix}
\bigskip

Here, we are going to derive the analytic continuation for ${\cal D}
(s,\nu,{\theta\over 2\pi})$ given by eq. (\ref{ca}), in the lines of
Ambjorn and Wolfran \cite{AW} and derive its main properties at
$s=0$. In fact their result corresponds to a particular case of ours,
when we take $\theta=0$.

Starting from eq.  (\ref{D}) and using the integral representation
for the Gamma function \cite{GR}

\begin{equation}
\Gamma(s) A^{-s} = \int_0^\infty d\tau \; \tau^{s-1} e^{-A\tau}\;,
\end{equation}

\noindent which is valid for $ Re\, \tau >0$ we find

\begin{eqnarray}
{\cal D} (s,\nu,{\theta\over 2\pi}) & = &  =
\sum_{m=-\infty}^{+\infty} \left[ \nu^2 +
\left({\theta\over 2\pi} + m \right)^2 \right]^{-s} \nonumber\\
& = & {1\over \Gamma (s)} \sum_{m=-\infty}^{+\infty} \int_0^\infty
dx\; x^{s-1} \exp\left\{ - \left[ \nu^2 +
\left({\theta\over 2\pi} + m \right)^2 \right] x \right\}\nonumber\\
& = & {1\over \Gamma (s)}  \int_0^\infty dx\; x^{s-1}  \exp\left\{-
\nu^2  x \right\}
\sum_{m=-\infty}^{+\infty}  \exp\left\{ -
\left({\theta\over 2\pi} + m \right)^2 x \right\}
\end{eqnarray}

Now, using the Poisson summation formula \cite{WW}

\begin{equation}
\sum_{n=-\infty}^{+\infty} e^{-(z+n)^2 {\pi\over\tau}}
= \sqrt{\tau} \sum_{n=-\infty}^{+\infty} e^{-n^2\pi\tau -2inz\pi}
\end{equation}

\noindent and identifying $m=n$, $z=\theta /2\pi$, $x=\pi /\tau$ we
can write

\begin{eqnarray}
{\cal D} (s,\nu,{\theta\over 2\pi}) & = & {1\over \Gamma (s)}
\int_0^\infty dx\; x^{s-1}  \exp\left\{- \nu^2  x \right\}
\sqrt{\pi\over x}
\sum_{m=-\infty}^{+\infty}  \exp\left\{ - m^2 {\pi^2\over x}
- i m \theta \right\} \nonumber\\ & = & {\sqrt{\pi} \over \Gamma (s)}
\sum_{m=-\infty}^{+\infty} e^{-im\theta} \int_0^\infty dx\;
x^{s-{1\over2}-1}  e^{- \nu^2  x  - m^2 {\pi^2\over x}} \;.
\end{eqnarray}

The integral in the last line of the above equation can be written in
terms of a modified Bessel function of the second kind, $K_\alpha
(y)$ \cite{GR}

\begin{equation}
\int_0^\infty
dx\; x^{\alpha-1}  e^{- \gamma x  - {\beta\over x}} = 2 \left(
{\beta\over\gamma} \right)^{\alpha\over 2} K_\alpha
(2\sqrt{\beta\gamma}) \;,
\end{equation}

\noindent which is valid for $ Re \beta\, >0$ and $ Re\, \gamma >0$.
Then, splitting the sum over $m$ from $-\infty$ to $+\infty$ into the
sums over negative values, $m=0$ and positive values and identifying
$\alpha=s-1/2$, $\beta=m^2\pi^2$ and $\gamma=\nu^2$, we have

\begin{equation}\label{DG}
{\cal D} (s,\nu,{\theta\over 2\pi}) =  {G(s)\over \Gamma (s)}\;,
\end{equation}

\noindent where we defined

\begin{equation}
G(s)= \sqrt{\pi}  \left\{ {\Gamma (s-{1\over 2})\over
\nu^{2s-1}} + 4 \sum_{m=1}^{+\infty} \cos(m\theta) \left(
{m\pi\over\nu}\right)^{s-{1\over 2}} K_{s-{1\over 2}}
(2m\pi\nu)\right\} \;,
\end{equation}

\noindent which is the result expressed in eq. (\ref{ca}).

Now, we are going to compute ${\cal D} (0,\nu,{\theta\over 2\pi})$ as
well as $\partial {\cal D} (0,\nu,{\theta\over 2\pi})/ \partial s$.
Noting that $G(s)$ is an entirely analytical function and that
$\Gamma (s)$ diverges as $s\rightarrow 0$, we see that

\begin{equation}\label{D0}
{\cal D} (0,\nu,{\theta\over 2\pi}) = 0 \;.
\end{equation}

\noindent
Next, we turn to the derivative of the zeta-function $\zeta^\theta
(s;L)$ at the origin,

\begin{eqnarray}
\left. {\partial\over\partial s} \zeta^\theta (s;L) \right|_{s=0}
& = & \left. {\partial\over\partial s} \left\{
\left({2\pi\over\beta}\right)^{-2s} {\cal D} (s,\nu, {\theta\over
2\pi}) \right\}\right|_{s=0}\nonumber\\ & = & \left.
{\partial\over\partial s} {\cal D} (s,\nu,{\theta\over 2\pi})
\right|_{s=0}\;,
\end{eqnarray}

\noindent where we already used the fact that
${\cal D} (0,\nu,{\theta\over 2\pi}) = 0 $.  Using eq.(\ref{DG}) we
have

\begin{equation}
\left. {\partial\over\partial s}
{\cal D} (s,\nu,{\theta\over 2\pi}) \right|_{s=0} = \left. - \left[
{\Gamma'(s) \over \Gamma^2 (s)} G(s) \right]
\right|_{s=0}
+\left. \left[{ 1 \over \Gamma (s)} G'(s) \right]
\right|_{s=0}
\end{equation}

\noindent and the second term of the r.h.s. of the above expression
vanishes too.  Noting that

$$\lim_{s\rightarrow 0} {\Gamma'(s) \over \Gamma^2 (s)} = - 1 $$

\noindent and using the fact that $\Gamma (-{1\over 2}) = -2
\sqrt{\pi}$, we have

\begin{eqnarray}
\left. {\partial\over\partial s}
{\cal D} (s,\nu,{\theta\over 2\pi}) \right|_{s=0} & = & G(0)
\nonumber\\
& = &  -2 \pi \nu  + 4 \sum_{m=1}^{+\infty} \cos(m\theta) \left(
{\nu\over m}\right)^{1\over 2} K_{-{1\over 2}} (2m\pi\nu) \;.
\end{eqnarray}

\noindent
Then, using the well known expression for $K_{-{1\over 2}} (z)$,
namely, \cite{GR}

\begin{equation}
K_{-{1\over 2}} (z) = \sqrt{\pi\over 2z} e^{-z}
\end{equation}

\noindent and the result

\begin{equation}
\sum_{n=1}^{\infty} {1\over n} e^{-2\pi\nu n} = \pi\nu - \ln [2 \sinh
(\pi\nu)]\;,
\end{equation}

\noindent we finally obtain

\begin{eqnarray}
\left. {\partial\over\partial s}
{\cal D} (s,\nu,{\theta\over 2\pi}) \right|_{s=0} & = &  -2 \pi \nu
+ 2 \sum_{m=1}^{+\infty} \cos(m\theta) {1\over m}
e^{-2m\pi\nu}\nonumber\\ & = &  -2 \pi \nu  +  \sum_{m=1}^{+\infty}
{1\over m} e^{-(2\pi\nu-i\theta)m}+  \sum_{m=1}^{+\infty} {1\over m}
e^{-(2\pi\nu+i\theta)m}\nonumber\\ & = & - \ln \left[ 2 \sinh \left(
\pi\nu + {i\theta\over 2}\right)\right] \nonumber\\ & = & - \ln
\left[ 2 \left( \cosh (2\pi\nu) -\cos \theta
\right)\right] \;,
\end{eqnarray}

\noindent which is precisely the result shown in eq.(\ref{dD}).

\vfill\eject

\end{document}